# REALTIME MONITORING FOR THE NEXT GENERATION OF RADIOTELESCOPES

**David G. Barnes[1], Grenville Armitage[2]**

[1] *Centre for Astrophysics and Supercomputing, Swinburne University of Technology, Mail number H39, PO Box 218, Hawthorn, VIC 3122, AUSTRALIA; david.g.barnes@gmail.com*

[2] *Centre for Advanced Internet Architectures, Swinburne University of Technology, Mail number H39, PO Box 218, Hawthorn, VIC 3122, AUSTRALIA; garmitage@swin.edu.au*

**ABSTRACT**

The forthcoming generation of radiotelescopes pose new and substantial challenges in terms of system monitoring. Information regarding environmental conditions, signal connectivity and level, processor utilisation, memory use, network traffic and even power consumption needs to be collected, displayed in realtime, and preserved in a permanent database. In this paper, we put forward the Ganglia monitoring system as a scalable, robust and efficient architecture that appears well-suited to the data collection aspect of radiotelescope monitoring, and we discuss approaches to the visual display of the streaming metric data produced by Ganglia. In particular, we present initial work in the use of 3-dimensional (3-d) multiplayer game technology for instantaneous status monitoring and enquiry, and we describe the extensions to this work required for radiotelescope monitoring.

**INTRODUCTION**

A number of major new radiotelescopes are currently in their final design phase or are even in the early stages of construction. Examples include the Murchison Widefield Array Low Frequency Demonstrator (MWA-LFD [1]), the Australian Square Kilometre Array Pathfinder (ASKAP) and the Karoo Array Telescope (meerKAT). These new facilities are typically being built to address very specific science questions, or they are intended as a demonstration of, and test facility for, new technologies.

The new challenges of the forthcoming generation of radiotelescopes can be appreciated using a simple parameterisation of telescope design, where $N$ is the number of collecting elements (traditionally parabolic dishes), $n$ is the number of signal detectors operating concurrently per collecting element, and $D$ is the size (typically diameter) of an individual element. The next generation radiotelescopes have small element sizes ($D = 5 – 15$m) but element and detector numbers greatly increased over their predecessors, best characterised by the product ($N \times n$) = 300 – 3000. Some designs (eg. MeerKAT) achieve large ($N \times n$) products by building hundreds of small dishes each with only one or two signal feeds. Other designs (eg. ASKAP) will achieve large ($N \times n$) products by building only ~30 dishes, but placing a focal plane array of ~100 detectors at the focus of each element.

Beyond their system complexity driven by the increased element and detector numbers, the new radiotelescopes are also typically distributed over a geographical area ~10km in diameter, with additional *remote stations* anywhere from 100km to 3000km distant. Different stations of the same telescope may measure signals that have propagated through independent parts of the Earth's ionosphere, leading to challenges in calibration and instrument stability. They may be on different power circuits, experience localised weather conditions or be subject to different operating restrictions.

Radiotelescopes are advanced and sophisticated devices that must be monitored for two main reasons:
- System integrity: is the telescope operating correctly and safely, and are all subsystems operating within their design range?
- Signal quality: is the telescope delivering "science-quality" data? That is, images, spectra or baseband datastreams that approach the theoretical capabilities of the telescope.

Monitoring data is used downstream from the collection point to provide a *passive, realtime* overview of the system status to telescope operators and engineers, and astronomers using the telescope. Subsets of the monitoring data may also be examined by *active, realtime* software that notes critical errors and automatically instigates a remedy or patch (eg. shuts down the effected components) and/or escalates the warning to a telescope operator or engineer.

In this paper, we discuss the collation, management and especially the display of monitoring data (MD) from the next-generation radiotelescopes (NGRTs).[1] The dramatically increased complexity (especially through total feed count $N$ x $n$), geographic spread of the stations and the general isolation of the NGRTs from population centres contribute new difficulties in realtime monitoring. Traditionally, MD volumes have been small enough that no serious effort at information management was required. Similarly, the relatively low total bandwidth of MD meant that quite simple visual displays sufficed to convey the entire system status to observers. With this in mind, we present the Ganglia Monitoring System (used for computer clusters) as an architecture that might be suitably adapted to the data collation and management operations for NGRTs. We then present a novel approach for displaying the MD inside a 3-dimensional game-like environment, and discuss the unique aspects of this method. We also discuss alternative display techniques for MD from NGRTs.

**THE GANGLIA MONITORING SYSTEM**

Ganglia is a monitoring system originally designed for computer clusters [2]. The core of Ganglia is a network-aware daemon (gmond) that is run on each node (computer) of a cluster. The gmond program regularly samples the state of its host (eg. CPU utilisation, network packet in/out rates, etc.) and multicasts (or unicasts) the measurements ("metrics") to the local network. Only metrics that have changed are emitted, thereby improving efficiency. The source node for the metrics in any particular multicast message are simply identified by the source IP address. While other monitoring systems (eg. Nagios[2]) provide a *different* daemon to (actively or passively) *collect* the metrics from a cluster of nodes, Ganglia's gmond program *also* fulfils this role by listening to multicasts on the local network. Accordingly, every gmond program on single cluster knows the current status of all nodes on the cluster. In the Ganglia architecture, there is no (implicit) master monitoring location: the cluster state can be obtained from any node. Furthermore there is no master list of nodes to monitor, so nodes can join and leave the monitoring network at will, without causing interruption to data collection or monitoring processes. The instantaneous cluster status can be obtained in XML format by simply connecting to any gmond program on a specified port.

Ganglia has been extended to handle "grids" of clusters [3]. This is accomplished with the gmetad program, which collects information from specified hosts that themselves are running gmetad or gmond programs. Gmetad differs to gmond in that it actively polls a specified set of hosts for information, rather than listening for broadcasts. This does require some explicit configuration of the monitoring network, however it does enable the hierarchical aggregation (hereafter *federation*) of MD from multiple clusters and clusters of clusters that do not share organisation-level networks.

Two end-user interfaces to the Ganglia MD are provided with Ganglia: gstat – a command-line tool – and gmetad plus a PHP-based web site. Gstat is a simple utility that can be used to print out the load and CPU state of each node in the monitoring network. Beyond collecting and federating MD, gmetad maintains a local, historical "round-robin database" (RRD) of the metrics. The RRD does not grow with time: as metrics age, they are averaged into longer and longer time periods. So while instantaneous information is available at full time-resolution (eg. 30s intervals), year-old data is only available at perhaps weekly time-resolution. The Ganglia web pages provide an elegant interface to the RRD, and can generate time-based plots (with time baselines from minutes to years) of all numeric metrics for individual nodes, for clusters, and for federated clusters.

**Ganglia for Radiotelescopes**

Ganglia is an interesting system to consider for NGRT monitoring for several reasons.
1. There is a natural mapping of the topology of a federated cluster of computers to the archetype NGRT:
   - an individual computer and its metrics map to an individual telescope element and its receiver(s);
   - a cluster of computers maps to a node or station, which comprises of order 10 telescope elements; and
   - a federated cluster maps to an entire radiotelescope, with tens to hundreds of stations (nodes) aggregated into one entity.

---

[1] Please note that in this paper, we are not interested in the mechanics of individual metric generation (eg. how the temperature at a particular geographical location is actually measured or digitised). This is a detail that is largely inconsequential in the context of the management and display of the MD.
[2] Nagios: http://www.nagios.org

2. Not only does the archetype NGRT have a high-performance cluster of $10^3 – 10^4$ CPUs as its *back end* processor (that itself should to be monitored), computers feature increasingly up the chain towards the *front end* receivers and digital samplers: an NGRT *is* a federated cluster of computers with specialised input devices.
3. Ganglia provides a mature, robust and efficient wide-area network monitoring architecture that is ideal for deployment across a geographically spread NGRT, and/or one isolated from large populations centres and high-bandwidth networks.

With these in mind, we now describe two approaches to applying the Ganglia Monitoring System to an archetype NGRT: a "quick and dirty" method, and a more structured and extensible strategy which entails some modifications to Ganglia.

**Quick-n-dirty Ganglia for NGRTs.** The standard installation of Ganglia includes a program called gmetric. This program, run on an individual computer in the monitoring network, can be used to add an arbitrary measurement to the set of metrics broadcast by that computer's local gmond program. For example, the command:
```
gmetric –n humidity –v 75 –t float –u % –d 300
```
adds a metric called 'humidity' to the set broadcast by the node's gmond, with a value of 75%. The lifetime of the metric is 300s, and after this time has elapsed, gmond will drop the metric. Shortly after issuing this command, the metric will be appear in the aggregated cluster data reported by any gmond or gmetad in the monitored network. Non-volatile data (eg. the latitude and longitude of a station) can be added once using a lifetime of 0s: these metrics never expire.

It is straightforward to imagine how the gmetric program might be used by a computer controlling one or more front-end systems of a NGRT. For each of the measurements it obtains, it makes a call to gmetric, providing a useful identifying name for the metric and the measured value and unit. Different measurements might be polled at different rates and inserted with different lifetimes. For cases where the metrics quickly become unwieldy[3] the gmetric program's *spoofing* feature can be used: using the `–S` argument, data that appears to come from other IP addresses can be inserted in the local gmond's set of metrics. These metrics then propagate onwards to the rest of the monitoring network. The spoofed IP addresses can be real (and may even match the addresses of other computers in the Ganglia network) or fake (and do not even need to be valid) and be associated with an arbitrary name. The spoofing feature can be used to (i) avoid sequential naming of metrics where one computer monitors several similar devices and which otherwise necessitates additional downstream parsing, and (ii) aggregate different measurements from an individual device (that might be sourced from different computers in the Ganglia network[4]) into the one logical unit.

Using gmetric to add structured metrics to the Ganglia monitoring network is a simple and relatively elegant operation. However, it is not without flaws: the gmetad program does not store the *ad hoc* metrics in its round-robin database, nor does it know the provenance or "meaning" of these metrics as it does for the *inbuilt* Ganglia metrics. Because the user-inserted metrics are not stored in the RRD, the web front-end cannot produce plots of them, and it simply displays the ephemeral values in text form. Despite these shortcomings, gmetric provides a quick way to extend Ganglia with a few specific measurements, especially if a custom front end is being used to display or monitor the data.

**Modified Ganglia for NGRTs.** With some relatively minor changes to address the issues identified in the "quick-n-dirty" approach above, Ganglia can be turned into an integrated system that provides instantaneous and historical monitoring data for the next generation of radiotelescopes. Specifically, the following two modifications would be sufficient:
1. rather than using gmetric, the standard set of metrics monitored by the gmond program is expanded to include measurements specific to the radiotelescope; and
2. the gmetad program is altered to store the expanded set of metrics to the RRD, and the Ganglia web pages are altered to include graphs of the additional metrics.

Every measurement added to the Ganglia MD must still be injected by a real computer running gmond, but judicious use of Ganglia's IP spoofing and cluster of clusters native support would be an ideal way to produce a hierarchical set of MD for a radiotelescope. The one, federated monitoring network would include:

---

[3] For example, a front-end computer which is the source of cryogenic temperature measurements for tens or hundreds of feeds may need to distinguish the measurements with sequence metric names like 'cryotemp1A', 'cryotemp1B', 'cryotemp2A', …
[4] For example, the cryogenic temperature of a receiver may be sampled by a different computer to the one that samples the power supply to that receiver.

- real, named computers, with standard performance and status metrics (CPU usage, load, disk free, core temperatures, etc.);
- real, named instruments (with forged or real IP addresses), with specific performance and status metrics (signal level, voltage supply, configuration, cryogenic temperature, vacuum strength, etc.); and
- named stations, incorporating one or more real computers and one or more real instruments.

Beyond this, metrics for back-end compute clusters, network switches and even external connectivity could easily be added to the federated monitoring data.

**THE HUMAN-COMPUTER INTERFACE FOR RADIOTELESCOPE MONITORING**

Radiotelescope monitoring data collected by Ganglia can be used in two principal ways: background software agents can monitor the real-time data and actively alert other subsystems or humans to existing or emerging problems; and foreground display software can use the data to present the instantaneous and historical status of the radiotelescope for visual inspection. In this paper we are concerned with the latter, and while the default Ganglia web interface could immediately provide a useful and flexible "human-computer inferface" for the NGRT MD, several alternatives are possible.

A key challenge for real-time data monitoring is concisely displaying the current state of multiple real-world devices, particularly when each device state may be represented by many independent metrics. For a typical NGRT, with the product ($N$ x $n$) in the vicinity of 1000, and of order 5 – 10 independent metrics per input device, there may be 10,000 or more independent measurements available instantaneously. While this number is small compared to the number of independent elements on a typical display (~1M pixels), it is well beyond the capacity of iconic and visual short-term memory [4]. Accordingly, a key design decision for the display of MD should be to only draw attention to change, and to metrics whose values are (or are soon likely to be) outside the acceptable operating range. A normal state across the entire NGRT system should be reflected by a relatively quiescent display of the MD. Except in rare cases, non-volatile metrics (for example, the specifications of a device) should not clutter the display of the MD. These data can be made available in a "drill-down" mode in the visualisation of monitoring data, but not in the top-level monitor display.

The use of a 3-dimensional (3-d) approach to display real-time monitoring data is largely unexplored to our knowledge, certainly for radiotelescopes. By 3-d, we mean any technique that constructs graphical representations of metrics in a 3-d space; it need not necessarily entail stereoscopic projection or immersive display. Our interest in using a 3-d space for visual monitoring of radiotelescopes is motivated by two key features of the monitoring data:
1. each element of data is geographically-tagged – it was recorded at a particular position in the real 3-d world; and
2. individual sets of metrics within the whole data set are aggregated into entities that reflect a real division of components in the system.

In addition, a key characteristic of data visualisation is that 3-d visualisation of abstract data is potentially more powerful than 2-d visualisation [5]. Apart from the self-evident feature that using a 3-d space for data visualisation immediately provides an additional ordinal axis for metric mapping, we suggest that a 3-d space can allow a more intuitive aggregation and grouping of metrics than the alternative of dividing a 2-d display into rectangular sections. The use of a 3-d space for monitoring is not without pitfalls: for example, it is easy to fall into the trap of producing something that "looks cool", but has little practical merit. There is also the danger that a user might select a viewpoint that obscures other important information or hides the "big picture"; for critical monitoring systems it is important to be aware of this and ensure the user can always see an overview, eg. in a small "always on top" panel.

A number of generalities in data visualisation should be considered in display monitoring data for the NGRTs, irrespective of whether a 2-d or 3-d approach is being used. For brevity we just list a few of them here; see [6] for excellent coverage on the general topic of effective graphical communication of data:
- where possible, visually encode the most important data using preattentive attributes such as form:orientation, form:size and color:hue (eg. [7]);
- avoid using the color:hue attribute for encoding quantitative attributes where quantitative visual comparison is important; and
- be consistent in attribute use: if one "page" of the visualisation uses attribute Y (eg. form:curvature) for metric X (eg. phase delay), then all pages displaying metric X should do so using attribute Y.

We now discuss a novel, 3-d approach to displaying monitoring data harvested using a Ganglia network.

**Leveraging 3D Game Engines (L3DGE) for Radiotelescope Monitoring**

Our approach is to construct a 3-d virtual world within which each monitored device is represented by an in-world entity whose observable attributes (such as spin rate or colour) are tied to the monitored real-world metrics. The observer exists and moves around within this virtual world, viewing entities from different angles and positions as they see fit. The virtual world itself, and the spatial placement of entities within the virtual world, is constructed to emphasise the relationship between the entities being watched and the real-world devices they represent. Mapping each different metric to *visually orthogonal*[5] entity behaviours ensures the viewer can see, and comprehend, the states of multiple metrics concurrently.

A specific example that looks promising for NGRT monitoring is L3DGEWorld,[6] a free and open-source tool kit with origins in IP network monitoring and control [8]. L3DGEWorld utilises off-the-shelf multiplayer 3-d game engine technology to create a virtual world within which arbitrary entities jump, spin, change size and colour in response to changes in one or more metrics being monitored in the real world. Unlike many 3-d tools for data representation, L3DGEWorld's focus is on concurrent representation of time-varying state 'as it is happening', to one or more observers, while allowing each observer to independently vary their in-world viewpoint over time as they see fit. Furthermore, L3DGEWorld allows observers to interact with in-world entities, optionally triggering events back in the real world.

L3DGEWorld may be used wherever there is a suitable source of monitored data (such as a Ganglia-equipped NGRT) and an appropriately designed virtual world. A proof-of-concept was released in August 2007 as LCMON 1.0 (L3DGEWorld Cluster MONitor) – a variant of L3DGEWorld 2.1 customised to monitor CPU load, memory consumption and network traffic rates across a 145-node, Ganglia-equipped supercomputer cluster. A 3-d 'star' floating in the virtual world represents each cluster node. Every 60 seconds LCMON polls the clusters' Ganglia monitoring infrastructure and updates the spin rate, size and bounce rate of each star. The concurrent visual representation of three distinct metrics for each node is remarkably effective. An observer can quickly gain a qualitative appreciation for the distribution of computational load across the cluster. Each cluster node's star spins faster with higher CPU load, is 'fatter' with higher memory consumption, and bounces at a rate proportional to the level of inbound network traffic.

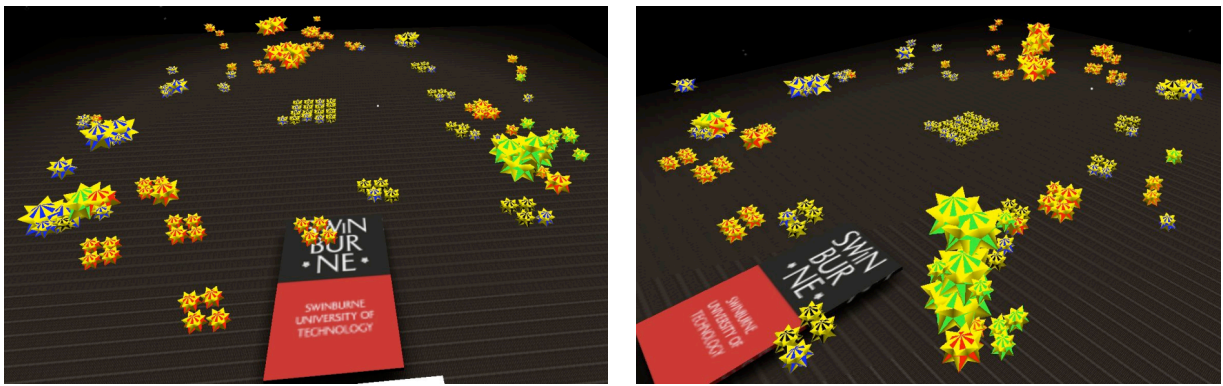

Fig. 1. L3DGEWorld being used to monitor the (synthetic) state of the (simulated) 32T component of the MWA LFD radiotelescope.

Fig. 1 illustrates how an NGRT – in this case, the 32T component of the MWA LFD – might appear within L3DGEWorld. Individual tiles of the 32T system are represented by four stars, and these groups of stars are clustered together and arranged spatially in the virtual world to represent the final 32T array layout (D. Oberoi, *priv. comm.*, 2007). The monitored state of each telescope tile is mapped into visually orthogonal, dynamic characteristics of the stars. For example, spin rate might relate to signal phase, diameter to signal amplitude, and bounce rate to delay. The colour attribute might be used to indicate status (eg. green: normal; amber: non-critical problem; red: fault or off-line), or mapped to another attribute such as equipment temperature. A core of 17 stars is added to the centre of the virtual world, which might be used to monitor the state of the on-line computer system and back-end processor for the array. L3DGEWorld allows an entity's shape to be changed on the fly, enabling stars to be replaced by other shapes or figures (for example, to visually distinguish particularly noteworthy changes in a telescope's monitored state, or in the case of a full-scale NGRT, perhaps indicate different observing projects using different components of the telescope).

---

[5] For example, an entity's spin rate and colour are visually orthogonal – under reasonable conditions an observer can easily differentiate spin rate and colour. However, spin rate on two different axes is likely to be an ambiguous way of representing two different metrics.
[6] http://caia.swin.edu.au/urp/l3dge

L3DGEWorld utilises a client-server architecture based on the Quake III Arena multiplayer game engine. The virtual world is instantiated by a L3DGEWorld server, a modified Quake III Arena engine that maintains virtual-world details by, for example, polling Ganglia-equipped nodes. Multiple observers (using modified Quake III Arena game clients) may connect to the L3DGEWorld server at the same time and navigate the virtual world, with each one seeing the state of the monitored system from their own perspective (as shown in Figures 1 and 2). Observers may be situated anywhere on the planet, requiring only a functional IP service between each observer's L3DGEWorld client and the NGRT's L3DGEWorld server.

**REMARKS AND CONCLUSION**

We have demonstrated a novel approach to monitoring real-time performance data from the archetype next generation radiotelescope. Our skeletal implementation uses the Ganglia monitoring system for metric aggregation and management, and the L3DGEWorld toolkit for displaying the metrics in an immersive, 3-d environment. Tools already exist to insert arbitrary volatile and non-volatile metrics into the monitoring network. The capability to forge data sources makes it possible to monitor non-standard devices and enables flexible aggregation of data. Accordingly, a complete implementation of the proposed system across a real NGRT, such as the MWA Low Frequency Demonstrator currently being deployed in Western Australia, is mostly trivial. The prospects for effectively using the multiplayer and interaction paradigms that the L3DGEWorld system inherits from its game engine foundations are good.

Using L3DGEWorld is just one way to embed and present Ganglia monitoring data in a 3-d virtual space. Another approach that is less immersive and gamelike (and therefore more traditional) but might nevertheless be useful, is to use a 3-d graphics programming library to construct a 3-d representation of the monitoring network and then allow the user to manipulate the virtual camera to obtain a meaningful projection of the system status. We have commenced work on a Python-based system that fetches the Ganglia monitoring data in XML format, parses it, and then constructs an S2PLOT [9] visualisation of the system state. Compared to the game engine approach to 3-d graphics, the S2PLOT approach lacks a straightforward mechanism for interacting with other "players" in the monitoring environment and the benefits that this might offer. However, S2PLOT does provide an extensive set of graphical primitives and high-level data display functions (such as volume rendering and isosurfaces) that might turn out to be very powerful for conveying the status of the large and complicated next generation radiotelescopes.

**ACKNOWLEDGEMENTS**

We thank Chris Fluke and Nick Jones for valuable discussions and ideas on data monitoring and visualisation, and Warren Harrop, Lucas Parry and Carl Javier for the development of L3DGEWorld and LCMON.